\documentclass[pdflatex,sn-mathphys-num]{sn-jnl}

\usepackage{graphicx}%
\usepackage{multirow}%
\usepackage{amsmath,amssymb,amsfonts}%
\usepackage{amsthm}%
\usepackage{mathrsfs}%
\usepackage[title]{appendix}%
\usepackage{xcolor}%
\usepackage{textcomp}%
\usepackage{manyfoot}%
\usepackage{booktabs}%
\usepackage{algorithm}%
\usepackage{algorithmicx}%
\usepackage{algpseudocode}%
\usepackage{listings}%

\usepackage{array}
\usepackage{threeparttable}

\usepackage{amsmath}
\usepackage{float}

\theoremstyle{thmstyleone}%
\theoremstyle{thmstyletwo}%

\theoremstyle{thmstylethree}%

\begin{document}

\title[Article Title]{\textbf{An Immune Infiltration-Based Risk Scoring System for Prognostic Stratification in Colorectal Adenocarcinoma}}


\author[1]{\fnm{Oluwafemi} \sur{Ogundare}}\email{femiogundare001@gmail.com}

\affil[1]{\orgdiv{Department of Medicine and Surgery}, \orgname{University of Ibadan}, \city{Ibadan}, \country{Nigeria}}

\abstract{\textbf{Background:} Colorectal adenocarcinoma (CRC) remains a leading cause of cancer-related mortality worldwide, with variable patient outcomes despite treatment advances. Traditional prognostic methods based on clinicopathological variables alone do not fully capture the biological complexity of the disease. This study aims to develop a risk scoring system based on genes associated with tumor-infiltrating immune cells (TIIC-associated genes) to improve prognostic assessment in CRC.

\textbf{Methods:} RNA-seq gene expression and clinicopathological data from TCGA-CRC (647 tumor samples, 51 normal tissues) were analyzed to identify differentially expressed TIIC-associated genes through comparison with the CIBERSORTx database. Univariate and multivariate Cox analyses were performed to screen for prognostic markers. A Gaussian mixture model was applied to cluster prognostic models and select the model with the most robust gene combination. The resulting risk scoring system was validated in an external cohort (GSE39582) and integrated with clinicopathological variables to develop a prognostic nomogram.

\textbf{Results:} From 128 TIIC-associated genes, an optimal prognostic model comprising CCL8 and TYR was identified. The risk score was calculated as 0.152×\textit{Exp}(CCL8)–0.516×\textit{Exp}(TYR). Kaplan-Meier analysis confirmed significant survival differences between high-risk and low-risk groups in both TCGA-CRC ({\unboldmath $p < 0.05$}) and GSE39582 ({\unboldmath $p < 0.05$}). Time-dependent ROC analysis showed AUC values ranging from 0.605 to 0.696 for 1-year, 3-year, and 5-year survival in TCGA-CRC and GSE39582. Multivariate Cox analysis identified TNM\_T, TNM\_N, and risk score as independent prognostic factors.

\textbf{Conclusion:} Our risk scoring system based on CCL8 and TYR effectively stratifies CRC patients into distinct prognostic groups and could guide treatment decisions, particularly when integrated with TNM staging in a nomogram.}




\maketitle

\section{Introduction}\label{sec1}
Colorectal adenocarcinoma (CRC) is a malignant tumor arising from the epithelial lining of the colon and rectum and represents the most common histological subtype of colorectal cancer [1]. It remains one of the leading causes of cancer-related mortality worldwide, with a 5-year survival rate ranging from 91\% for early-stage disease to 14\% for metastatic disease [2]. Although advancements in surgery, chemotherapy, immunotherapy, and targeted therapies have expanded and improved treatment options, patient outcomes remain highly variable [3]. 

Immune infiltration is central to the tumor microenvironment, influencing tumor progression, metastasis, and treatment response [4]. Studies have shown that genes associated with tumor-infiltrating immune cells (TIIC-associated genes) drive tumorigenesis and determine clinical outcomes in CRC [5, 6]. Thus, there is a need for prognostic models that incorporate immune-related molecular features to stratify patients and guide treatment decisions. 

Traditional prognostic assessments rely on clinicopathological variables such as tumor size, lymph node involvement, and histological grade, which, while informative, do not fully capture the biological complexity of the disease [7]. Gene expression-based prognostic models offer a more targeted approach by leveraging transcriptomic data to predict patient outcomes [8]. Using mathematical modeling techniques such as Cox regression and Gaussian mixture modeling, these models can identify key prognostic genes and stratify patients into distinct risk groups [9]. 

In this study, we developed a risk scoring system based on TIIC-associated genes to improve prognostic assessment in CRC. Using transcriptomic data from the TCGA-CRC cohort, we identified differentially expressed TIIC-associated genes and screened for prognostic markers using univariate and multivariate Cox regression analyses. A Gaussian mixture model (GMM) was applied to cluster prognostic models and select the most robust gene combination. This approach identified an optimal prognostic model comprising CCL8 and TYR, which was used to construct a risk scoring system. Kaplan-Meier survival analysis confirmed that the risk scoring system effectively stratifies patients into high- and low-risk groups, with significant survival differences.

\section{Materials and methods}\label{sec2}
\subsection{Acquisition of TIIC-associated genes}\label{subsec2}
A dataset of TIIC-associated genes was obtained from the CIBERSORTx database (\url{https://cibersortx.stanford.edu/}), which provides a leukocyte gene signature matrix (LM22) consisting of 547 genes across 22 immune cell types. These include natural killer cells, T cells, naïve B cells, memory B cells, plasma cells, monocytes, macrophages, dendritic cells, mast cells, eosinophils, and neutrophils.
\subsection{Data collection and processing}\label{subsec2}
RNA-seq gene expression and clinicopathological data for 647 CRC solid tumor samples and 51 normal tissue samples (TCGA-CRC) were obtained from the Pan-Cancer Atlas 
(\url{https://portal.gdc.cancer.gov/}). The sequencing reads were mapped to the GRCh38 human genome assembly, with each sample containing expression profiles for 60,660 genes. TCGA-CRC count data was normalized using variance stabilizing transformation to ensure uniform variance across genes with different expression levels. For external validation, the GSE39582 dataset was retrieved from the GEO database 
(\url{https://www.ncbi.nlm.nih.gov/geo/}). Normalization of GSE39582 was performed using the robust multi-array average algorithm, followed by batch effect correction using the ComBat algorithm.
\subsection{Identification of differentially expressed genes (DEGs) and TIIC-associated genes in TCGA-CRC cohort}\label{subsec2}
DEGs between tumor and normal tissue samples in TCGA-CRC were identified using DESeq2, with an adjusted $p < 0.05$ and a $\log \text{ fold change} > 1.5$
. A Venn diagram was used to determine the overlap between DEGs in TCGA-CRC and the 547 TIIC-associated genes. A total of 128 TIIC-associated genes was identified for downstream analysis. Gene expression patterns of the 128 genes were visualized using a heatmap generated with the pheatmap package (version 1.0.12) in R. Functional enrichment analysis was performed using the clusterProfiler (version 4.0), org.Hs.eg.db (version 3.5.0), and GOplot (version 1.0.2) packages in R.
\subsection{Screening for prognosis related TIIC-associated genes}\label{subsec2}
Univariate Cox analysis was performed on the 128 TIIC-associated genes to screen for the most significant prognostic genes, using $p < 0.05$.
\subsection{Multivariate Cox analysis and GMM}\label{subsec2}
The selected prognostic genes were then used to construct multiple prognostic models using multivariate Cox analysis. Various combinations of these genes were tested, and the resulting models were clustered using GMM, a probabilistic approach that represents data as a combination of multiple Gaussian distributions. The model with the highest AUC score within any of the clusters was selected as the optimal prognostic model.
\subsection{Construction of risk scoring system}\label{subsec2}
A risk scoring system was constructed using the genes from the optimal prognostic model. The risk score was calculated as a weighted sum of the expression levels of these genes, with their respective coefficients from the multivariate Cox analysis serving as weights.
\subsection{Evaluation of risk scoring system}\label{subsec2}
Patients in the TCGA-CRC cohort were stratified into high-risk and low-risk groups based on the optimal Kaplan-Meier cutoff of the risk score. The predictive performance of the risk scoring system was assessed using time-dependent receiver operating characteristic (ROC) curves at 1-year, 3-year, and 5-year intervals. This analysis was conducted using the tROC package (version 0.4) in R.
\subsection{External validation}\label{subsec2}
The risk scoring system was also externally validated on GSE39582 to further evaluate its performance.
\subsection{Identification of independent prognostic factors}\label{subsec2}
Univariate and multivariate Cox analyses were performed on clinicopathological variables, including gender, TNM staging (TNM\_T, TNM\_N, TNM\_M), and the risk score to evaluate their association with survival outcomes and determine whether they can serve as independent prognostic factors.
\subsection{Construction and verification of a nomogram}\label{subsec2}
All independent prognostic factors identified through multivariate Cox analysis were used to construct a nomogram for predicting survival, which was evaluated using calibration analysis and decision curve analysis.

\section{Results}\label{sec2}
\subsection{Identification of TIIC-associated genes in TCGA-CRC cohort}\label{subsec2}
Using DESeq2, a total of 6,011 DEGs were identified between 647 CRC tumor samples and 51 normal tissue samples in TCGA-CRC, based on an adjusted $p < 0.05$ and a $\log \text{ fold change} > 1.5$. A Venn diagram analysis showed 128 overlapping TIIC-associated genes between the 6,011 DEGs and the 547 TIIC-associated genes obtained from the CIBERSORTx database, which were subsequently used for downstream analysis (Fig. 1A). To validate these findings, the expression patterns of the 128 TIIC-associated genes were examined in the external dataset GSE39582 (Fig. 1B). Functional enrichment analysis was then performed to explore the biological significance of these genes. KEGG pathway enrichment analysis showed significant enrichment in pathways such as cytokine-cytokine receptor interaction, chemokine signaling, viral protein interaction with cytokines and receptors, hematopoietic cell lineage, and B cell receptor signaling (Fig. 1C). Gene Ontology (GO) analysis showed that processes related to cell killing, humoral immune response, and chemokine-mediated signaling were significantly enriched (Fig. 1D).

\begin{figure}[H] 
    \centering
    \includegraphics[width=0.8\textwidth]{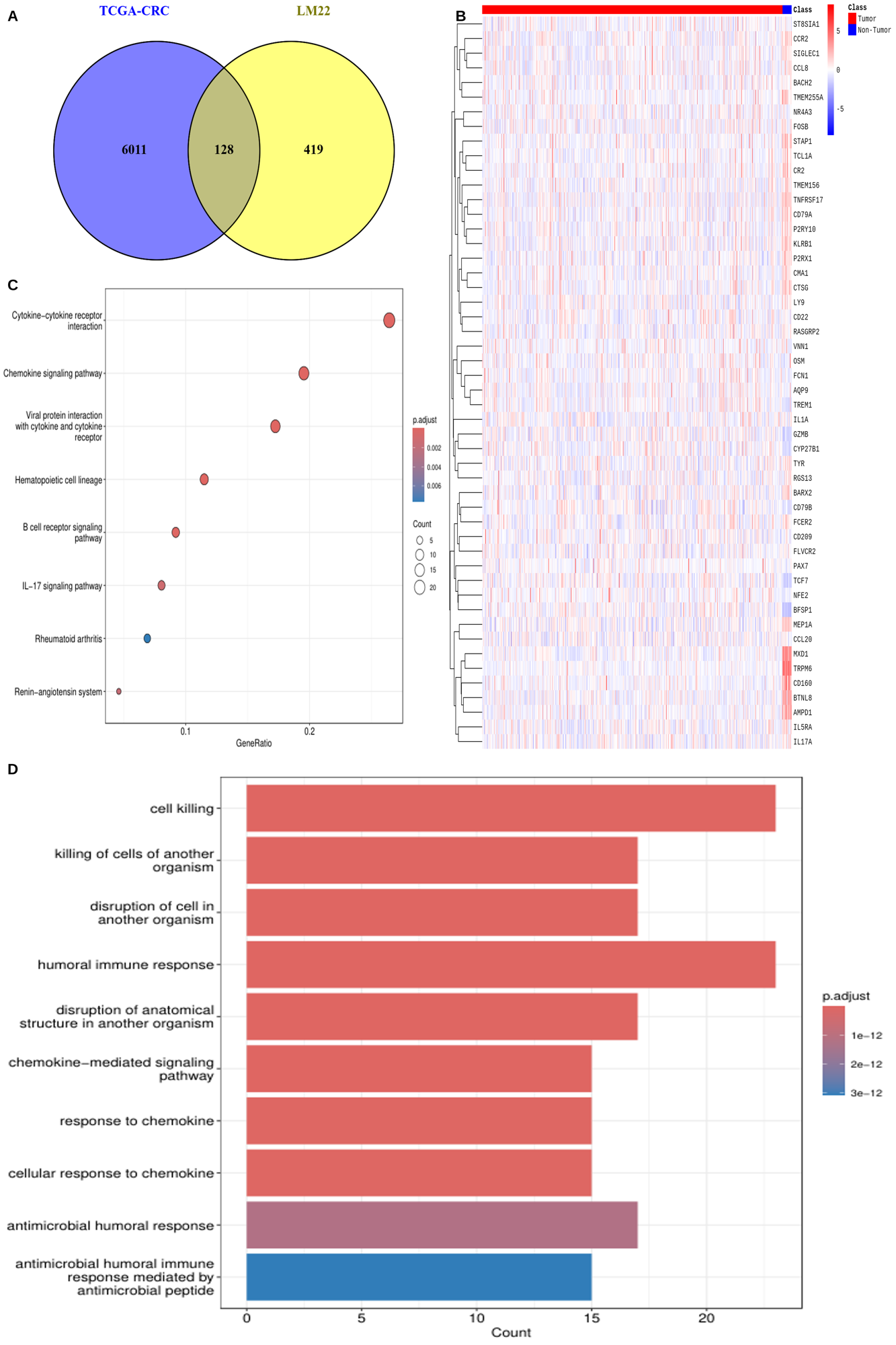}
    \caption{Identification and functional enrichment analysis of TIIC-associated genes in TCGA-CRC cohort. \textbf{A} Venn diagram showing the intersection between DEGs and TIIC-associated genes. \textbf{B} Heatmap of first 50 of 128 TIIC-associated genes in GSE39582. \textbf{C} Gene ontology analysis. \textbf{D} KEGG enrichment analysis.}
    \label{fig:figure1} 
\end{figure}

\subsection{Identification of prognosis related TIIC-associated genes}\label{subsec2}
Univariate Cox analysis identified 7 candidate prognostic genes (Table 1), which were selected for risk score modeling.
\vspace{1\baselineskip}

\renewcommand{\arraystretch}{1.3} 

\noindent
\textbf{Table 1:} Seven prognostic TIIC-associated genes identified by univariate Cox analysis.\par
\vspace{0.5em}

\noindent
\begin{tabular}{|p{2cm}|p{6cm}|p{4cm}|}
\hline
\textbf{Gene} & \textbf{Full Name} & \textbf{HR (95\% CI)} \\
\hline
GNG7     & G Protein Subunit Gamma 7              & 1.266 (1.059 -- 1.515) \\
\hline
SIGLEC1  & Sialic Acid Binding Ig Like Lectin 1   & 1.170 (1.026 -- 1.334) \\
\hline
RASGRP2  & RAS Guanyl Releasing Protein 2         & 1.245 (1.020 -- 1.520) \\
\hline
CXCL3    & C-X-C Motif Chemokine Ligand 3         & 0.866 (0.756 -- 0.992) \\
\hline
BACH2    & BTB And CNC Homology 2                 & 1.247 (1.011 -- 1.537) \\
\hline
CCL8     & C-C Motif Chemokine Ligand 8           & 1.175 (1.006 -- 1.371) \\
\hline
TYR      & Tyrosinase                             & 0.587 (0.349 -- 0.987) \\
\hline
\end{tabular}

\vspace{0.5em}
\noindent
\textit{Annotation:} HR = Hazard Ratio; CI = Confidence Interval. Values derived from univariate Cox regression analysis.

\subsection{Multivariate Cox analysis and GMM}\label{subsec2}
A multivariate Cox analysis was performed on various combinations of the 7 prognostic TIIC-associated genes, generating a total of 127 prognostic models. These models were then clustered using GMM, which grouped them into 5 distinct clusters based on their characteristics. The optimal prognostic model was identified in cluster 3 and achieved the highest AUC score. This model included two genes, CCL8 and TYR. The coefficients and hazard ratios of this model are presented in Table 2.

\vspace{1\baselineskip}

\renewcommand{\arraystretch}{1.3} 

\noindent
\textbf{Table 2:} The coefficient and hazard ratios of CCL8 and TYR in the optimal multivariate Cox model.\par
\vspace{0.5em}

\noindent
\begin{tabular}{|p{1.5cm}|p{2.0cm}|p{2.6cm}|p{2.6cm}|p{2.8cm}|}
\hline
\textbf{Gene} & \textbf{Coefficient} & \textbf{Hazard Ratio} & \textbf{Lower 95\% CI} & \textbf{Upper 95\% CI} \\
\hline
CCL8 & 0.152 & 1.164 & 0.998 & 1.359 \\
\hline
TYR  & -0.516 & 0.597 & 0.355 & 1.003 \\
\hline
\end{tabular}

\vspace{0.5em}
\noindent
\textit{Annotation:} CI = Confidence Interval. Values derived from multivariate Cox regression analysis.

\subsection{Construction of the risk scoring system}\label{subsec2}
A risk scoring system for predicting the prognosis of CRC patients was constructed based on the expressions of CCL8 and TYR. Using the coefficients from the multivariate Cox analysis, the risk score was calculated as follows:
\begin{equation}
\mathrm{Risk\ score} = 0.152*\,\mathrm{Exp}(\mathit{CCL8}) - 0.516*\,\mathrm{Exp}(\mathit{TYR}).\label{eq1}
\end{equation}
where,
\begin{align}
\mathrm{Exp} = \text{Expression value}
\end{align}

\subsection{Evaluation of risk scoring system}\label{subsec2}
Patients in the TCGA-CRC cohort were stratified into high-risk and low-risk groups based on the optimal Kaplan-Meier cutoff of the risk score. Figure 2A illustrates the relationships between risk scores, patient survival times, and the expression levels of CCL8 and TYR. The risk plot suggests that high-risk patients generally exhibit shorter survival times and poorer prognoses compared to low-risk patients. Also, higher CCL8 expression is associated with increased risk scores, while higher TYR expression is associated with lower risk scores. Figure 2B shows a statistically significant difference in survival between high-risk and low-risk patients ($p < 0.05$), with low-risk patients demonstrating higher survival probabilities and longer overall survival times. Figure 2C presents the time-dependent ROC curves, with AUC values of 0.607, 0.605, and 0.619 for 1-year, 3-year, and 5-year survival, respectively.

\begin{figure}[H] 
    \centering
    \includegraphics[width=0.8\textwidth]{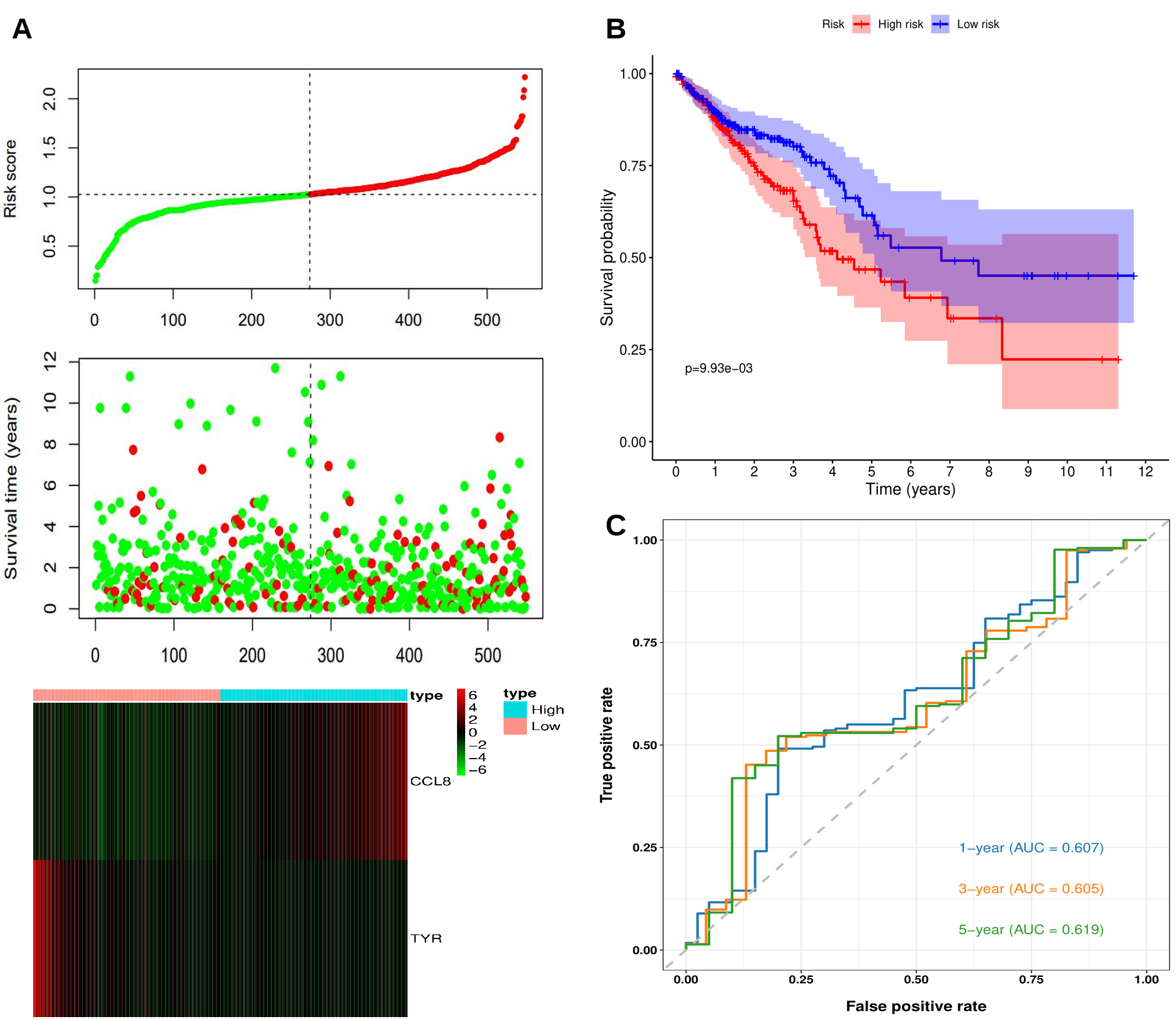}
    \caption{Risk score analysis, survival analysis, and prognostic performance of the risk scoring system in TCGA-CRC. \textbf{A} Risk scores, survival times, and expression levels of CCL8 and TYR. \textbf{B} Kaplan-Meier analysis of overall survival between high-risk and low-risk groups. \textbf{C} Time-dependent ROC curves for 1-year, 3-year, and 5-year survival.}
    \label{fig:figure2} 
\end{figure}

\subsection{External validation}\label{subsec2}
GSE39582 (566 tumor samples, 19 normal tissues) was used for external validation of the risk scoring system. Batch effects from technical variations (e.g., sequencing platforms, sample processing, RNA extraction methods, reagent lots, storage conditions, and operator handling) were corrected using the Combat algorithm, with TCGA-CRC as the reference. Before correction (Figure 3A), PCA plots showed distinct separation due to batch effects. After correction (Figure 3B), variance was reduced, and the datasets followed a more similar distribution.

\begin{figure}[H] 
    \centering
    \includegraphics[width=0.8\textwidth]{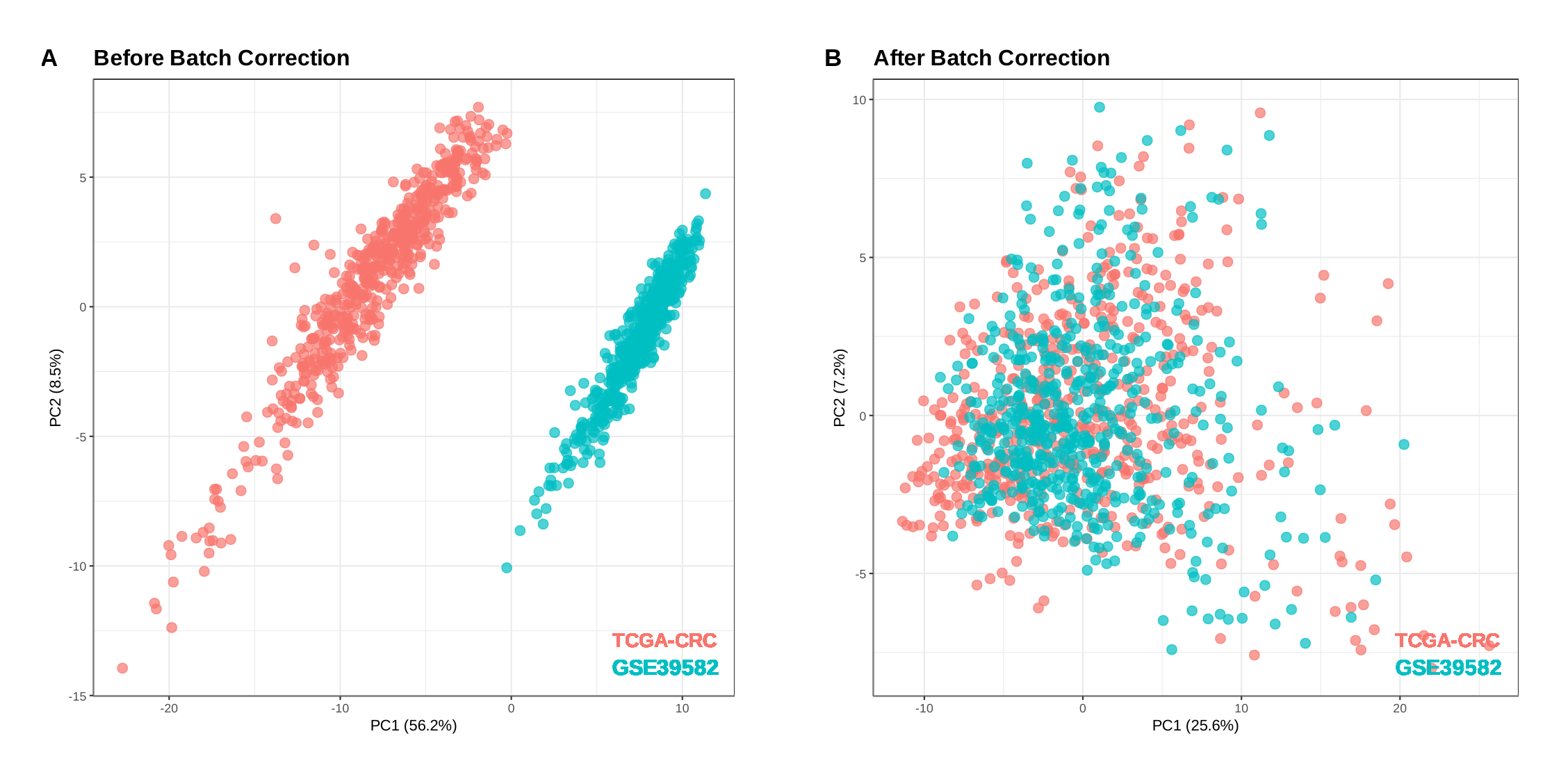}
    \caption{PCA plots for samples in TCGA-CRC and GSE39582. \textbf{A} Before batch correction. \textbf{B} After batch correction.}
    \label{fig:figure3} 
\end{figure}

Risk score was calculated for each patient using the same risk scoring formula. Patients were stratified into high-risk and low-risk groups based on the optimal Kaplan-Meier cutoff of the risk score. Figure 4A shows the relationships between risk scores, patient survival times, and the expression levels of CCL8 and TYR. The risk plot indicates that high-risk patients have shorter survival times and poorer prognoses compared to low-risk patients. Also, CCL8 expression is positively correlated with the risk score, whereas TYR expression is negatively correlated. Figure 4B confirmed a statistically significant survival difference between high-risk and low-risk patients ($p < 0.05$), with low-risk patients exhibiting higher survival probabilities and longer overall survival times. Figure 4C displays the time-dependent ROC curves, with AUC values of 0.632, 0.640, and 0.696 for 1-year, 3-year, and 5-year survival, respectively.

\begin{figure}[H] 
    \centering
    \includegraphics[width=0.8\textwidth]{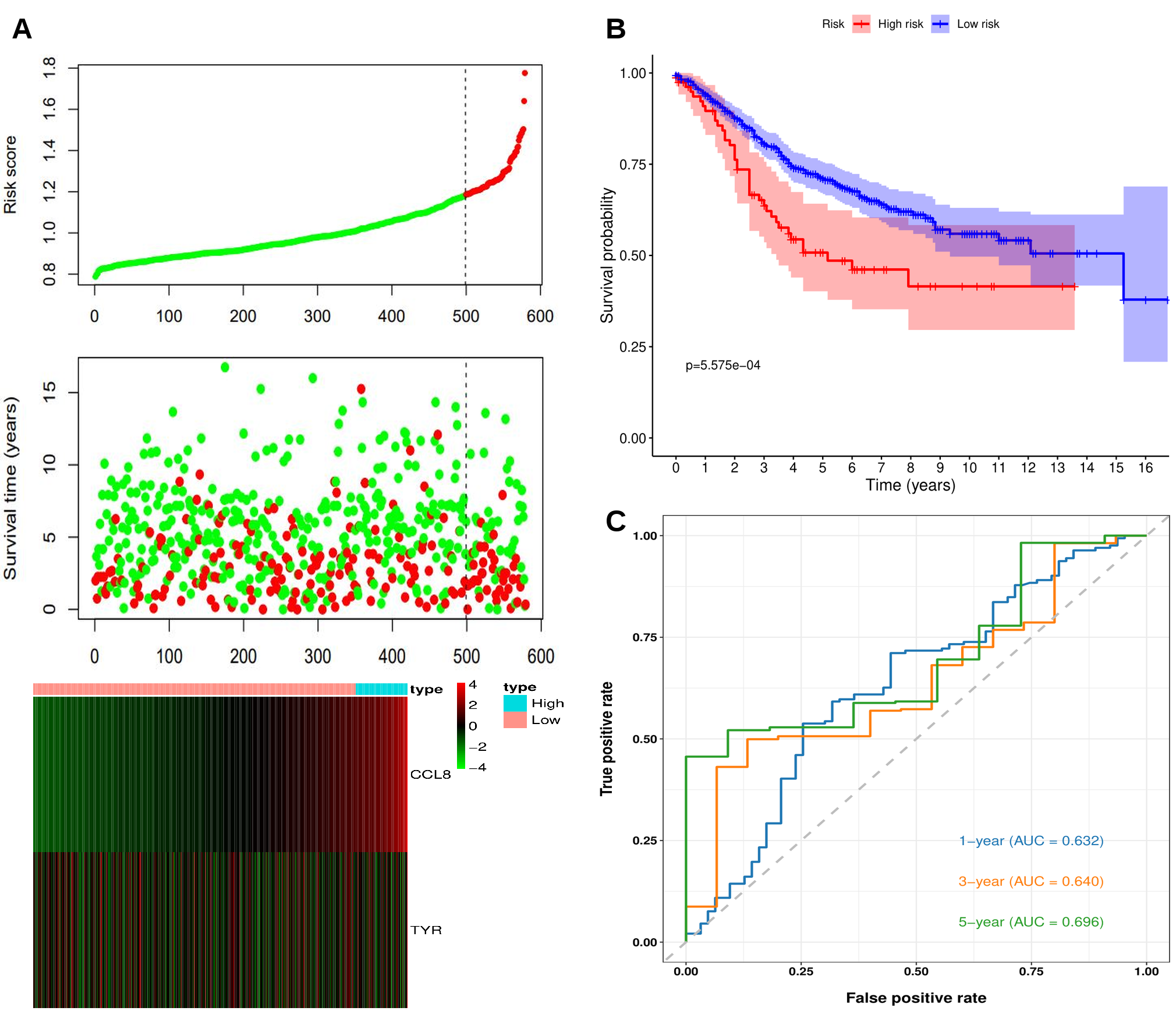}
    \caption{Risk score analysis, survival analysis, and prognostic performance of the risk scoring system in GSE39582. \textbf{A} Risk scores, survival times, and expression levels of CCL8 and TYR. \textbf{B} Kaplan-Meier analysis of overall survival between high-risk and low-risk groups. \textbf{C} Time-dependent ROC curves for 1-year, 3-year, and 5-year survival.}
    \label{fig:figure4} 
\end{figure}

\subsection{Identification of independent prognostic factors}\label{subsec2}
Univariate Cox analysis found TNM\_T, TNM\_N, TNM\_M, and risk score to be significantly associated with survival ($p < 0.05$), whereas gender was not (Fig. 5A). Multivariate Cox analysis identified TNM\_T, TNM\_N, and risk score as independent prognostic factors for CRC ($p < 0.05$), while TNM\_M was not independent and could be represented by other clinicopathological variables (Fig. 5B).

\begin{figure}[H] 
    \centering
    \includegraphics[width=0.8\textwidth]{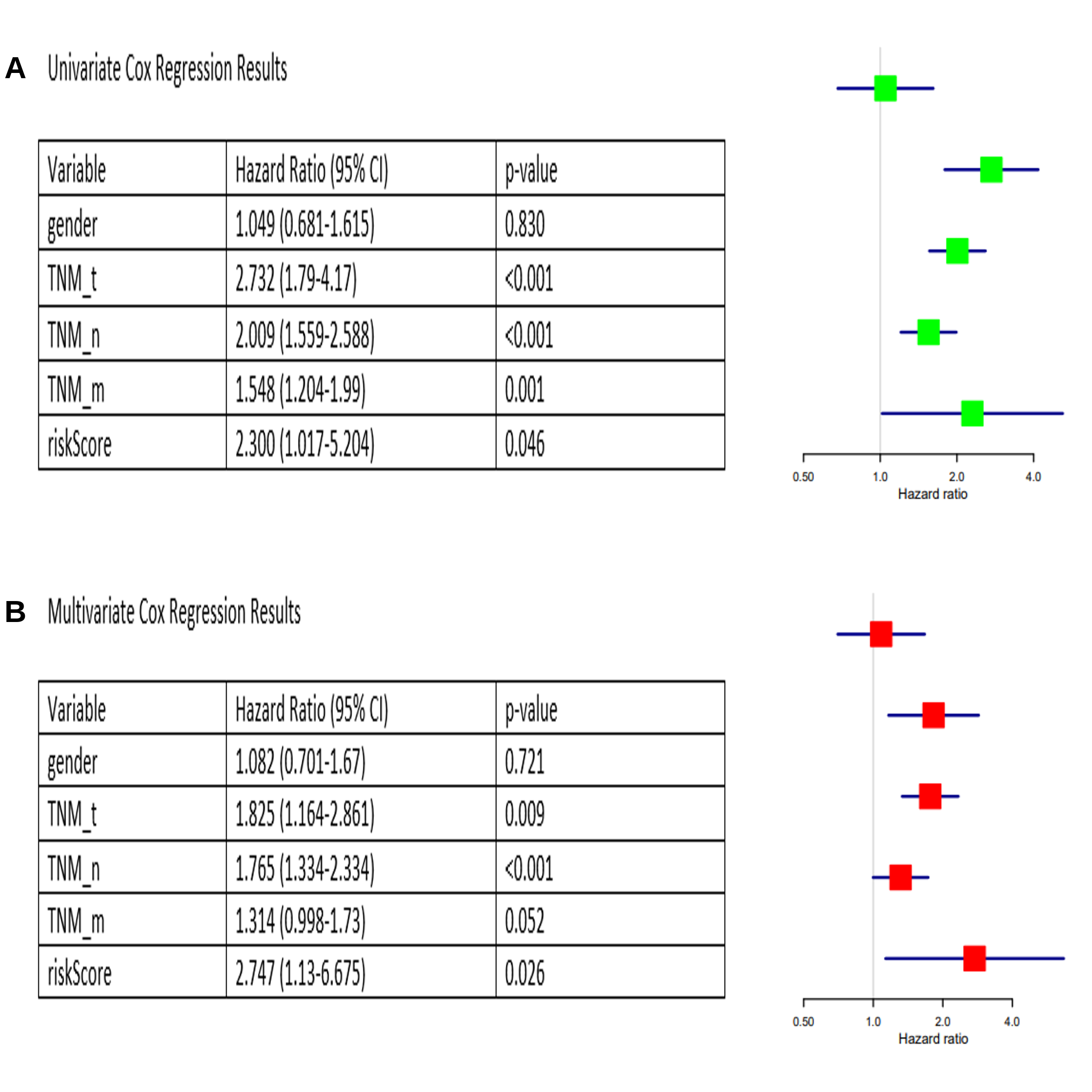}
    \caption{Univariate and multivariate Cox analyses for identification of independent prognostic factors. \textbf{A} Univariate Cox results. \textbf{B} Multivariate Cox results.}
    \label{fig:figure5} 
\end{figure}

\subsection{Construction and verification of a nomogram}\label{subsec2}
TNM\_T, TNM\_N, and risk score were incorporated into a nomogram (Fig. 6).
\begin{figure}[H] 
    \centering
    \includegraphics[width=0.8\textwidth]{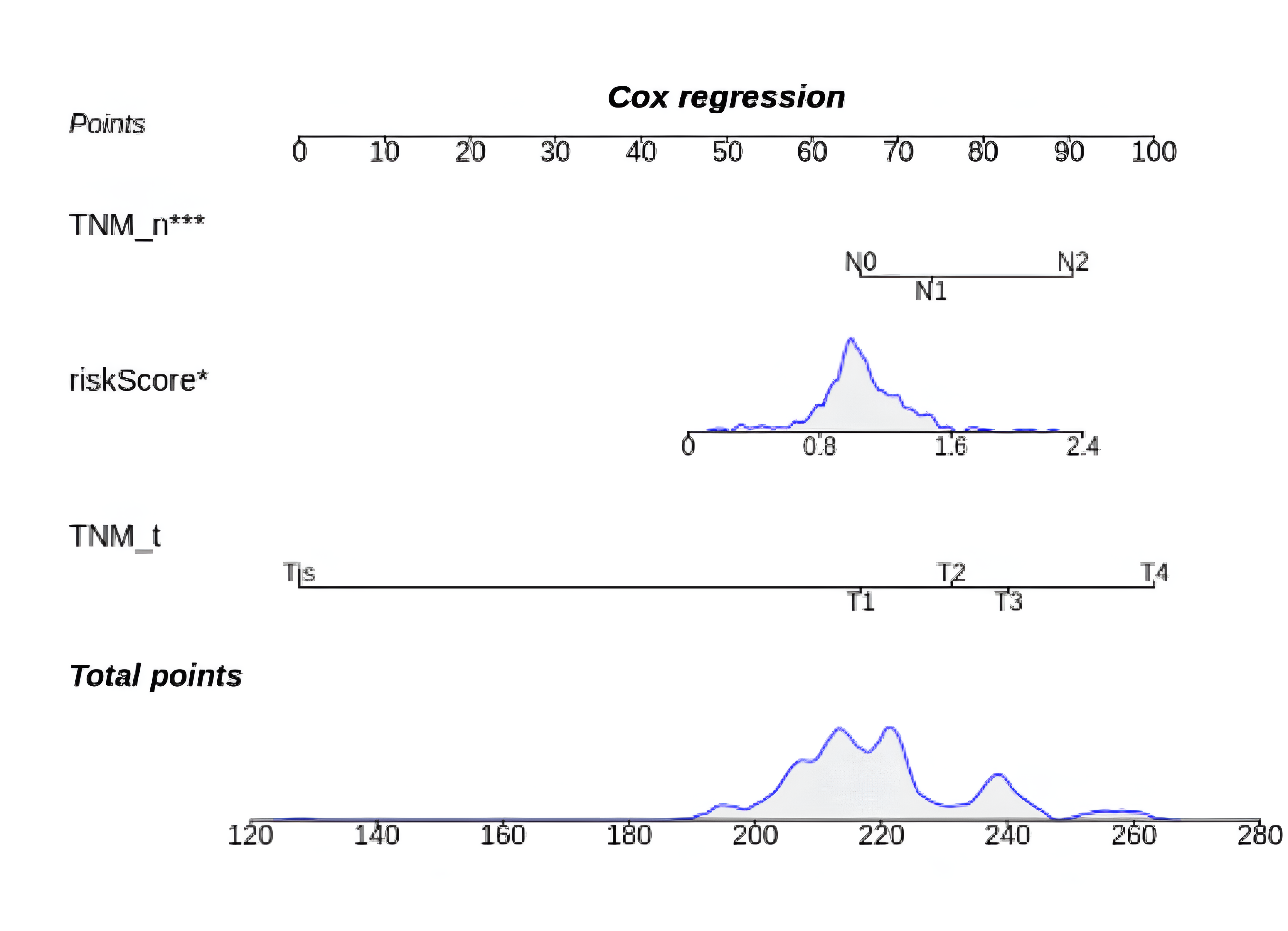}
    \caption{Prognostic nomogram. A nomogram integrating TNM\_T, TNM\_N, and risk score to estimate prognosis. Higher total points indicate a greater risk of poor survival outcomes.}
    \label{fig:figure6} 
\end{figure}
The 1-year, 3-year, and 5-year overall survival predictions of the nomogram were evaluated using calibration analysis (Fig. 7A, B, and C) and decision curve analysis (Fig. 7D, E, and F). The results showed that the nomogram demonstrated favorable prognostic performance.
\begin{figure}[H] 
    \centering
    \includegraphics[width=0.8\textwidth]{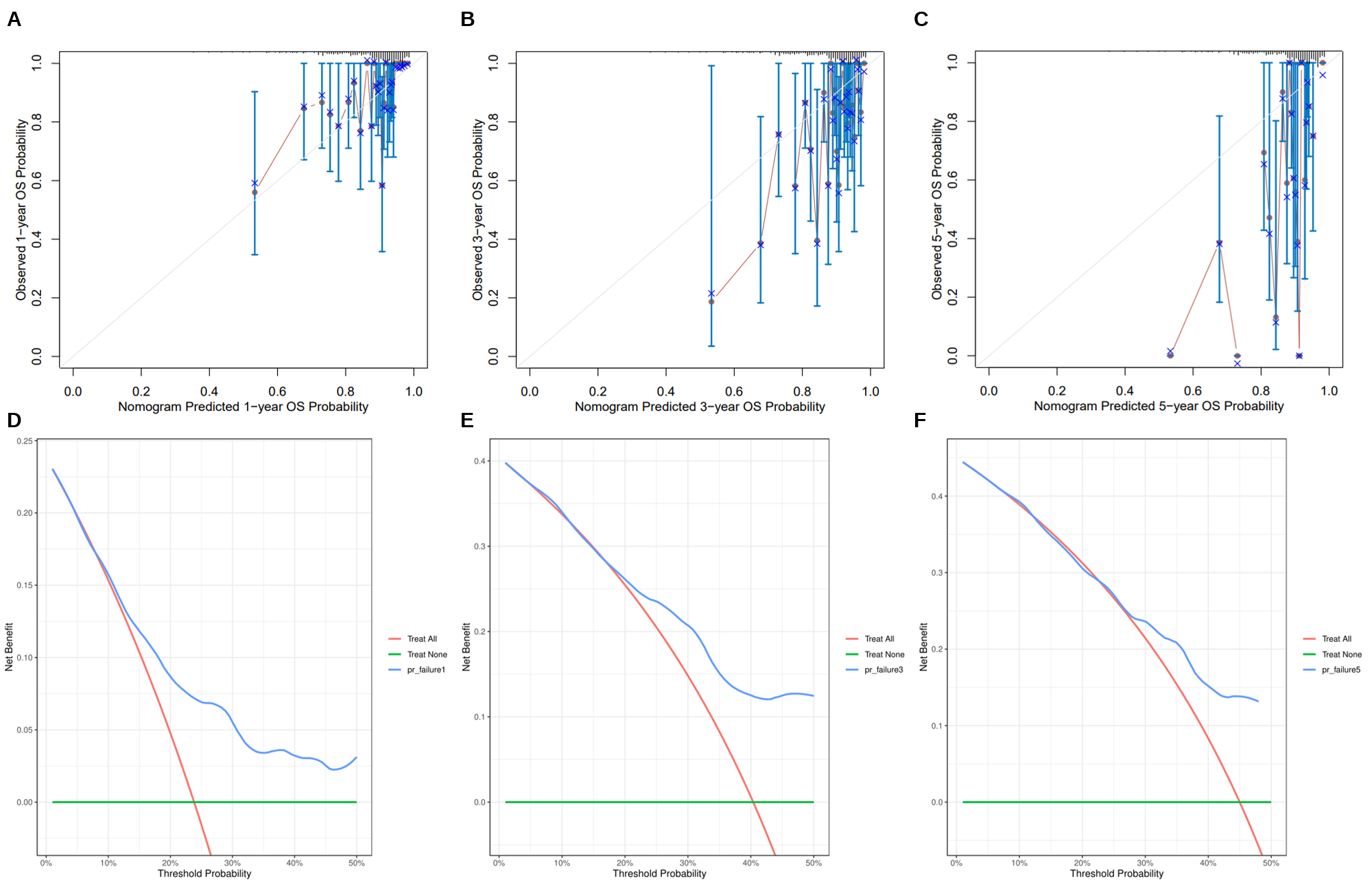}
    \caption{Calibration analysis and decision curve analysis of the nomogram. \textbf{(A–C)} Calibration curves for 1-year, 3-year, and 5-year survival. The x-axis represents the predicted overall survival probability, while the y-axis shows the actual observed survival. The 45-degree gray line represents the ideal prediction. \textbf{(D–F)} Decision curve analysis for 1-year, 3-year, and 5-year survival, showing the clinical net benefit at different threshold probabilities.}
    \label{fig:figure7} 
\end{figure}

\section{Discussion}\label{sec2}
In this study, we established a novel prognostic risk scoring system for CRC based on two TIIC-associated genes, CCL8 and TYR. Our findings demonstrate that this risk scoring system effectively stratifies patients into high-risk and low-risk groups with significantly different survival outcomes, both in the TCGA-CRC cohort and in GSE39582, the external validation dataset. We also showed that the risk score, together with TNM\_T and TNM\_N, serves as an independent prognostic factor, allowing for the construction of a nomogram with favorable prognostic performance.

The optimal prognostic model identified through our multivariate Cox analysis and GMM comprises two genes with opposing effects on patient prognosis, namely CCL8 and TYR. CCL8 (C-C Motif Chemokine Ligand 8) showed a positive coefficient of 0.152 in our model, indicating that higher expression is associated with increased risk and poorer outcomes. In contrast, TYR (Tyrosinase) exhibited a negative coefficient of -0.516, suggesting that elevated expression correlates with reduced risk and improved survival.

CCL8, also known as monocyte chemoattractant protein-2 (MCP-2), is a chemokine that attracts monocytes, lymphocytes, basophils, and eosinophils by interacting with several chemokine receptors [10]. Its role in cancer progression has been well documented. Studies have shown that CCL8 promotes breast cancer dissemination, enhances the migration and invasion of esophageal cancer cells, and drives lung cancer progression [11–14]. Interestingly, CCL8 has also been reported to inhibit melanoma metastasis, suggesting that its function may be highly variable across cancer types [15]. In colorectal cancer specifically, elevated CCL8 expression has been linked to increased tumor cell invasion and migration [16, 17]. Our findings align with these observations in epithelial cancers, as we found that higher CCL8 expression correlates with increased risk scores and poorer survival outcomes.

TYR encodes an enzyme that targets melanosomes in melanocytes and serves as a rate-limiting enzyme in melanin synthesis [18, 19]. While primarily known for its role in pigmentation, recent evidence suggests that TYR may also contribute to immune regulation and oxidative stress response [20, 21]. In our study, TYR expression was negatively correlated with risk scores, indicating a potential protective role in CRC. Interestingly, although TYR itself appears to have a tumor-suppressive effect, TYR kinase, which phosphorylates it, has been previously implicated in the development of adenomatous polyps, ulcerative colitis, and CRC [22]. This finding warrants further investigation into the mechanisms by which TYR might influence tumor suppression or immune surveillance in colorectal cancer.

Traditional prognostic assessment for CRC relies heavily on the TNM staging system, which, while informative, does not fully capture the biological heterogeneity of the disease [7]. Several gene expression-based prognostic signatures have been proposed for colorectal cancer, including the Oncotype DX Colon Cancer Assay (12-gene panel) and ColoPrint (18-gene panel) [23, 24]. However, these signatures often require larger gene sets and may not specifically incorporate immune-related genes.

Our risk scoring system uses only two genes, making it more cost-effective and clinically applicable. Clinicians can assess patient prognosis and make informed decisions by simply calculating the risk score. It is also based on TIIC-associated genes, which are involved in the tumor microenvironment and recognized as drivers of cancer progression [5, 6].

The risk scoring system may have meaningful clinical applications. By stratifying patients into high-risk and low-risk groups, it could guide treatment decisions. High-risk patients may benefit from more aggressive treatment and closer monitoring, while low-risk patients may be spared unnecessary interventions and their associated toxicities. Also, integrating our risk score with established clinicopathological variables (TNM\_T and TNM\_N) in a nomogram provides a valuable tool for personalized risk assessment. Beyond prognosis, CCL8 and TYR may also represent novel therapeutic targets in CRC.

Our study has several limitations. First, while the risk scoring system demonstrated significant prognostic value, the AUC values (ranging from 0.605 to 0.696) suggest that predictive accuracy could be improved. Future studies should explore integrating additional molecular features, such as DNA methylation or lncRNA expression, to improve the model performance. Second, our study was based on retrospective cohorts, which may have introduced selection bias. Prospective validation in clinical trials would provide stronger evidence for the clinical utility of our risk scoring system. In vivo and in vitro functional studies are also needed to better understand the precise mechanisms by which CCL8 and TYR influence CRC progression.
Moreover, while our study focused on overall survival, future research could assess the prognostic value of our model for other important clinical endpoints, such as disease-free survival, progression-free survival, or response to specific therapeutic regimens. Finally, the heterogeneity of colorectal adenocarcinoma, including differences between right-sided and left-sided tumors and various molecular subtypes (e.g., microsatellite instability status, CpG island methylator phenotype), was not fully addressed in our analysis. Investigating how our risk scoring system performs across these different subgroups could identify subtype-specific prognostic patterns.

\section{Conclusion}\label{sec2}
In summary, we have developed and validated a novel risk scoring system based on two TIIC-associated genes, CCL8 and TYR, that effectively stratifies CRC patients into distinct prognostic groups. The risk score, together with clinicopathological variables, serves as an independent prognostic factor and can be integrated into a nomogram for comprehensive risk assessment. Further validation is needed to evaluate the robustness of this system, while functional studies could clarify the role of CCL8 and TYR in CRC.

\section*{Declarations}
\begin{itemize}
\item \textbf{Author contribution:} \textbf{Oluwafemi Ogundare} is the sole contributor to this work and is responsible for all aspects of the study, including conception, design, analysis, interpretation, and manuscript preparation.
\item \textbf{Competing interests:} \textbf{Oluwafemi Ogundare} holds a patent unrelated to the subject matter of this manuscript. This does not influence the content or conclusions of the paper.
\item \textbf{Funding:} Not applicable.
\item \textbf{Ethics approval and consent to participate:} Not applicable.
\item \textbf{Consent for publication:} Not applicable.
\item \textbf{Data availability:} The datasets used in this study are publicly available in the TCGA (\url{https://portal.gdc.cancer.gov/}) and GEO (\url{https://www.ncbi.nlm.nih.gov/geo/}) databases.
\item \textbf{Author information:} \textbf{Oluwafemi Ogundare} is a final-year medical student in the Department of Medicine and Surgery, Faculty of Clinical Sciences, University of Ibadan, Nigeria.
\end{itemize}

\section*{References}
\begin{enumerate}
    \item Alzahrani SM, Al Doghaither HA, Al-Ghafar AB. General insight into cancer: An overview of colorectal cancer (review). Mol Clin Oncol. 2021;15:1–8.
    \item Siegel RL, Wagle NS, Cercek A, Smith RA, Jemal A. Colorectal cancer statistics, 2023. CA Cancer J Clin. 2023;73:233–54.
    \item Kumar A, Gautam V, Sandhu A, Rawat K, Sharma A, Saha L. Current and emerging therapeutic approaches for colorectal cancer: A comprehensive review. http://www.wjgnet.com/. 2023;15:495–519.
    \item Scott EN, Gocher AM, Workman CJ, Vignali DAA. Regulatory T Cells: Barriers of Immune Infiltration Into the Tumor Microenvironment. Front Immunol. 2021;12:702726.
    \item Ge P, Wang W, Li L, Zhang G, Gao Z, Tang Z, et al. Profiles of immune cell infiltration and immune-related genes in the tumor microenvironment of colorectal cancer. Biomedicine \& Pharmacotherapy. 2019;118:109228.
    \item Wu X, Li J, Zhang Y, Cheng Y, Wu Z, Zhan W, et al. Identification of immune cell infiltration landscape for predicting prognosis of colorectal cancer. Gastroenterol Rep (Oxf). 2022;11.
    \item Chang GJ, Hu CY, Eng C, Skibber JM, Rodriguez-Bigas MA. Practical application of a calculator for conditional survival in colon cancer. Journal of Clinical Oncology. 2009;27:5938–43.
    \item Stephenson AJ, Smith A, Kattan MW, Satagopan J, Reuter VE, Scardino PT, et al. Integration of gene expression profiling and clinical variables to predict prostate carcinoma recurrence after radical prostatectomy. Cancer. 2005;104:290–8.
    \item Zhang X, Yang L, Zhang D, Wang X, Bu X, Zhang X, et al. Prognostic assessment capability of a five-gene signature in pancreatic cancer: a machine learning based-study. BMC Gastroenterology 2023 23:1. 2023;23:1–10.
    \item Van Damme J, Struyf S, Opdenakker G. Chemokine–protease interactions in cancer. Semin Cancer Biol. 2004;14:201–8.
    \item Farmaki E, Chatzistamou I, Kaza V, Kiaris H. A CCL8 gradient drives breast cancer cell dissemination. Oncogene 2016 35:49. 2016;35:6309–18.
    \item Kim ES, Nam SM, Song HK, Lee S, Kim K, Lim HK, et al. CCL8 mediates crosstalk between endothelial colony forming cells and triple-negative breast cancer cells through IL-8, aggravating invasion and tumorigenicity. Oncogene 2021 40:18. 2021;40:3245–59.
    \item Zhou J, Zheng S, Liu T, Liu Q, Chen Y, Tan D, et al. MCP2 activates NF-kB signaling pathway promoting the migration and invasion of ESCC cells. Cell Biol Int. 2018;42:365–72.
    \item Yan C, Luo L, Urata Y, Goto S, Li TS. Nicaraven reduces cancer metastasis to irradiated lungs by decreasing CCL8 and macrophage recruitment. Cancer Lett. 2018;418:204–10.
    \item Yang P, Chen W, Xu H, Yang J, Jiang J, Jiang Y, et al. Correlation of CCL8 expression with immune cell infiltration of skin cutaneous melanoma: potential as a prognostic indicator and therapeutic pathway. Cancer Cell Int. 2021;21:1–11.
    \item Zhou H, Yao J, Zhong Z, Wei H, He Y, Li W, et al. Lactate-Induced CCL8 in Tumor-Associated Macrophages Accelerates the Progression of Colorectal Cancer through the CCL8/CCR5/mTORC1 Axis. Cancers (Basel). 2023;15:5795.
    \item Yamane T, Kanamori Y, Sawayama H, Yano H, Nita A, Ohta Y, et al. Iron accelerates Fusobacterium nucleatum–induced CCL8 expression in macrophages and is associated with colorectal cancer progression. JCI Insight. 2022;7.
    \item Shibahara S, Torruta Y, Sakakura T, Nager C, Chaudhuri B, Müller R. Cloning and expression of cDNA encoding mouse tyrosinase. Nucleic Acids Res. 1986;14:2413–27.
    \item Hearing VJ. Determination of Melanin Synthetic Pathways. Journal of Investigative Dermatology. 2011;131:E8–11.
    \item Zhang J, Yu R, Guo X, Zou Y, Chen S, Zhou K, et al. Identification of TYR, TYRP1, DCT and LARP7 as related biomarkers and immune infiltration characteristics of vitiligo via comprehensive strategies. Bioengineered. 2021;12:2214–27.
    \item Ipson BR, Fisher AL. Roles of the tyrosine isomers meta-tyrosine and ortho-tyrosine in oxidative stress. Ageing Res Rev. 2016;27:93–107.
    \item Malecka-Panas E, Kordek R, Biernat W, Tureaud J, Liberski PP, Majumdar AP. Differential activation of total and EGF receptor (EGF-R) tyrosine kinase (tyr-k) in the rectal mucosa in patients with adenomatous polyps, ulcerative colitis and colon cancer. Hepatogastroenterology. 1997;44:435–40.
    \item Bailey H, Turner M, Stoppler MC, Chao C. The 12-gene Oncotype DX Colon Recurrence Score (RS) test: Experience with > 20,000 stage 2 patients (pts). Journal of Clinical Oncology. 2018;36 4\_suppl:618–618.
    \item Kopetz S, Tabernero J, Rosenberg R, Jiang Z-Q, Moreno V, Bachleitner-Hofmann T, et al. Genomic Classifier ColoPrint Predicts Recurrence in Stage II Colorectal Cancer Patients More Accurately Than Clinical Factors. Oncologist. 2015;20:127–33.
\end{enumerate}

\end{document}